# Chasing the thermodynamical noise limit in whispering-gallery-mode resonators for ultrastable laser frequency stabilization

Jinkang Lim[1,*], Anatoliy A. Savchenkov[2], Elijah Dale[2], Wei Liang[2], Danny Eliyahu[2], Vladimir Ilchenko[2], Andrey B. Matsko[2,*], Lute Maleki[2], and Chee Wei Wong[1,*]

[1] *Mesoscopic Optics and Quantum Electronics Laboratory, University of California, Los Angeles, CA 90095.*

[2] *OEwaves Inc., 465 North Halstead Street, Suite 140, Pasadena, CA 91107.*

*Author e-mail addresses: jklim001@ucla.edu, andrey.matsko@oewaves.com, cheewei.wong@ucla.edu

**Ultrastable high-spectral-purity lasers have served as the cornerstone behind optical atomic clocks, quantum measurements, precision optical-microwave generation, high resolution optical spectroscopy and sensing. Hertz-level lasers stabilized to high finesse Fabry-Pérot mirror cavities are typically used for these studies but are large and fragile such that they have remained laboratory instruments. There is a clear demand in rugged miniaturized lasers operating potentially at comparable stabilities to those bulk lasers. Over the past decade, ultrahigh-$Q$ optical whispering-gallery-mode (WGM) resonators have served as a platform for low-noise microlasers but have not yet reached the ultimate stabilities defined by their fundamental noise. Here, we show the noise characteristics of WGM resonators and demonstrate a resonator-stabilized laser at the fundamental limit by compensating the intrinsic thermal expansion of a WGM resonator, allowing a sub-25 Hz linewidth and a 32 Hz Allan deviation on the 191 THz carrier in 100 ms integration. We also reveal the environmental sensitivities of the resonator at the thermodynamical noise limit and long-term frequency drifts governed by random-walk-noise statistics.**

High precision optical frequency metrology, spectroscopy [1,2], atomic clocks [3-5], optical interferometry [6], ultralow phase noise microwave generation [7], and light detection and ranging [8] benefit from stable and spectrally pure laser oscillators. Such low frequency noise oscillators can be achieved by stabilizing laser oscillators to a high quality factor cavity resonance ($Q = \nu/\Delta\nu$, where $\nu$ is the resonance frequency and $\Delta\nu$ is the full-width at half maximum of resonance). When the signal-to-noise ratio (SNR) of the detected laser signal is high enough in the measurement bandwidth (BW), the frequency stability can also be improved, which is scaled by $(Q \times SNR_{BW})^{-1}$. The benchmark ultrahigh-$Q$ resonances in nature are atomic transitions. The precise transitions of trapped atoms have been utilized for improving stability and frequency noise of both microwave and optical oscillators. For instance, the cesium hyperfine transition is the core building block of the well-developed microwave atomic clock and the optical transitions of trapped neutral and ionic atoms are used for modern optical atomic clocks. Although a laser with frequency instability of $6 \times 10^{-16}$ for 2 to 8 s integration time has been demonstrated in a cryostat via spectral-hole burning written in the $Eu^{3+}$:$Y_2SiO_2$ absorption spectrum [9], making small atomic traps is demanding such that the development of solid-state compact optical references is of immense interest. Laser oscillators with high-finesse optical Fabry-Pérot (FP) cavities have been demonstrated with sub-Hz linewidths and fractional frequency instabilities at the $10^{-15}$ levels in 1 to 10 s integration time, which is the noise limit imposed by the unavoidable thermal motion of the cavity's reflection multilayer coating, by either using ultra-low-expansion material in vacuum or temperature cooling to operate the cavity at the zero thermal expansion point in a cryostat [10-15]. However, direct miniaturization of such FP mirror cavities to a microscale size is challenging due to the quality of the cavity mirrors such that their applications have largely remained in the laboratory environment.

Over the last decade, ultrahigh-$Q$ WGM resonators [16-17] have been implemented for developing low-noise microlasers and microclocks [18-21], which show the broad transparent window and ultrahigh-$Q$ resonances without sophisticated dielectric mirror coatings and are tolerant to mechanical noise. Since the early studies on the fundamental thermal fluctuations in microspheres by Gorodetsky, M. *et al*. [22], theoretical predictions indicate that the thermodynamically bounded frequency instability of the WGM resonators can be better than $10^{-13}$ in 1 s integration time [23,24] if the proper material and stabilization technique are selected.

However, in spite of the tremendous progress, the existing WGM resonators still experience large frequency instabilities and long-term frequency drifts [25-28] hindering their use in precision metrology and timing applications.

Here, we show the noise characteristics of a WGM resonator at the fundamental themodynamical noise limit. We employ a thermal-compensation design for reducing the thermal sensitivity of a conventional WGM resonator and reveal the residual environmental sensitivities of the compensated resonator. Subsequently, we suppress the environmental perturbations using an evacuated rigid enclosure and demonstrate that a laser stabilized to the thermal-compensation WGM resonator, working as an optical probe, shows a spectral linewidth of less than 25 Hz and a fractional frequency instability of $1.67\times10^{-13}$ ($5.0\times10^{-12}$) on the 191 THz carrier at 0.1 (1) s integration time, which is the best among the WGM resonators of the given size and morphology without a stringent ambient temperature control. Furthermore, we confirm that the centre-shifted random walk noise statistics, imposed by the correlation between the WGM resonator temperature change and the ambient temperature or pressure variation, triggers the long-term frequency instability and monotonic frequency drift of the laser stabilized to the resonator at the thermodynamical noise limit.

**Results**

**Thermal-compensation ultrahigh-$Q$ WGM resonator**

The thermal-expansion coefficient of the crystalline $MgF_2$ is approximately 9 ppm $K^{-1}$, that is a large value compared with the conventional reference FP cavities possessing less than 0.1 ppm $K^{-1}$. To reduce the thermal sensitivity causing the thermo-mechanical fluctuations via expansion of the resonator length, we compensate the thermal sensitivity of a thin $MgF_2$ WGM resonator. The thermal compensation, sandwiching a WGM resonator with laminated Zerodur as illustrated in the inset of Figure 1a, is guided by numerical simulations and predicts the significant reduction of the thermal sensitivity (Supplementary Note 1). The design is applied to a $MgF_2$ WGM resonator possessing a 6.9 mm diameter and 100 (25) μm resonator (rim) thickness. The thickness of the Zerodur layer (up and down) used for the MgF$_2$ WGM resonator is 500 μm. To find the degree of thermal compensation, we take a pair of WGM resonators and mount them on the same temperature-stabilized platform. While two low-noise continuous wave (cw) lasers are locked to the WGM resonances respectively, the temperature of one of the WGM resonators is slowly changed and the relative frequency shift of the beat note between the two stabilized lasers is measured. We assume all slow changes in the beat frequency are attributed to the thermal expansion of the WGM resonator because the thermorefractive coefficient of crystalline MgF$_2$ is significantly smaller than the thermal expansion coefficient. The measurement in Figure 1a shows approximately 7 times improvement compared with a conventional MgF$_2$ WGM resonator. Although the enhancement factor is smaller than the value predicted by numerical simulations, attributed to the residual thermal expansion due to the imperfection in the device fabrication, the measurement confirms the validity of our thermal-compensation design. The compensated WGM resonator is packaged into a small form factor (40×40×15 mm$^3$) with an integrated thermal sensor and a Peltier-type thermoelectric cooler under the WGM resonator as well as a prism coupler and a photodetector. Laser light is delivered to the prism coupler using a polarization-maintaining single mode fiber with a firmly mounted output tip. The temperature of the WGM resonator is stabilized at 301.2 K by a proportional-integral-derivative (PID) feedback control using the thermoelectric cooler. We then measure the thermal sensitivity again by measuring the beat frequency between a cw laser stabilized to the WGM resonator and a cw laser referenced to the ultrastable FP cavity (Stable Laser Systems) possessing 1 Hz linewidth and 0.1 Hz s$^{-1}$ drift-rate while the set-temperature of the PID control changes, resulting in 3 ppm K$^{-1}$. From the measured thermal sensitivity, we calculate the noise limits imposed by the thermorefractive and thermal-expansion sensitivities of the resonator (Supplementary Note 2), which are the two dominant thermal noise sources for WGM resonators as illustrated in Figure 1b. After the thermal compensation, the thermal-expansion noise limit is lower than the thermorefractive noise limit near the carrier frequency as shown in Figure 1b such that the thermorefractive noise-limited fluctuations can be unveiled. The WGM resonator has the unloaded resonance bandwidth of 26 kHz and the loaded resonator $Q$ is characterized with 4 µs ring-down time corresponding to $2.4\times10^9$ at 191 THz carrier (Supplementary Note 2).

**Resonant frequency shift due to ambient perturbations**

The mechanism behind the resonant frequency shift of a WGM resonator by the ambient perturbation is the interaction between the evanescent wave of the WGM resonator and the air refractive index change caused by temperature and pressure variations. The temperature and pressure are, in principle, coupled quantities connected by the ideal gas law ($PV = nRT$) in a rigid box, where $P$ is the pressure, $V$ is the volume, $n$ is the number of moles, $R$ is the ideal gas constant, and $T$ is the absolute temperature. Therefore, the ambient temperature stability required for achieving the thermodynamical noise limited fluctuation can be derived from the pressure stability measurement. For this measurement we place the thermal-compensation WGM resonator in a rigid vacuum chamber on vibration isolation pads. A 3 kHz laser operating at 191 THz is stabilized to the compensated WGM resonator using the Pound-Drever-Hall (PDH) locking technique [29] that provides sufficient technical noise suppression and thus enforces the laser frequency to chase the resonance (Methods and Supplementary Note 3). The resonant frequency shifts induced by pressure changes are measured by counting the beat frequency between the stabilized laser and the FP reference laser at every second. Figure 2a shows the laser frequency shift when the pressure slowly increases ($\Delta P = 45$ mPa s$^{-1}$) in the vacuum chamber from $P_0 = 17$ Pa and 25 Pa respectively. At the given pressure $P_0$ and the change ($\Delta P$), we measure 13 (9) kHz with $P_0 = 17$ (25) Pa for a transverse magnetic (TM) mode input. Here, we define that the TM mode of our WGM resonator has the electric field distribution primarily in the radial direction. We also measure the resonant frequency shift at the different pressure increment. While the pressure in the vacuum chamber is increased from 17 Pa with $\Delta P = 130$ mPa s$^{-1}$, we measure the frequency shift of 36 kHz (Figure 2b), which shows approximately a linear relationship between the frequency shift and the speed of pressure change. To confirm our measurements, we quantify the impact of the refractive index change of the surrounding air medium on the WGM resonant frequency shift for TM and transverse electric (TE) modes using equations derived from the first order perturbation theory in ref [30].

$$\frac{\Delta v_{TM}}{v_0} = -\frac{\Delta n_{air}}{(n_{TM}^2 - 1)^{3/2}} \frac{\lambda}{2\pi r}, \qquad (1)$$

$$\frac{\Delta v_{TE}}{v_0} = -\frac{\Delta n_{air}}{(n_{TE}^2 - 1)^{3/2}} \left(2 - \frac{1}{n_{TE}^2}\right) \frac{\lambda}{2\pi r}, \qquad (2)$$

$$\Delta n_{air} = n_0 |\Delta P / P_0|, \qquad (3)$$

where $r$ is the radius of the WGM resonator, $n_{TE}$ and $n_{TM}$ are the refractive indices of the resonator host material, $\lambda$ is the wavelength, $\Delta n_{air}$ is the change of the refractive index of air and $n_0$ is the residual refractive index of air given by the air itself, that is $\sim 3\times 10^{-4}$, and therefore $\Delta n_{air} \approx 7.94\times 10^{-7}$ in these measurements. The calculated results from Eq. 1 for $P_0 = 17$ (25) Pa are $\Delta v_{TM} = 13.3$ (9.1) kHz at $\Delta P = 45$ mPa s$^{-1}$, which agrees well with our measured values of 13 (9) kHz for a TM mode input as illustrated in Figure 2a. The theoretical frequency shift from thermorefractive noise of our WGM resonator is estimated by $\langle(\Delta \omega_{tre})^2\rangle = \omega^2(\alpha_n T)^2 k_B (CV_m\rho)^{-1} \approx (2\pi\, 46\text{ Hz})^2$, where $\alpha_n$ is the thermorefractive coefficient, $T$ is the cavity temperature, $k_B$ is the Boltzmann constant, $C$ is the specific heat capacity, $V_m$ is the mode volume, and $\rho$ is the density of the host material. The values used for the theoretical estimates are given in Supplementary Note 2. Theoretical estimations and experimental measurements indicate that the pressure change ($\Delta P$) has to be controlled under 0.154 mPa to unveil the fundamental thermorefractive noise limit of the compensated WGM resonator. Using the ideal gas law, we calculate the corresponding temperature variation of 2.7 mK, in which $n = 3.53\times 10^{-6}$ mole, $V = 0.52$ L, and $R = 0.082$ atm·L mol$^{-1}$·K$^{-1}$ are used. This measurement shows that pressure and temperature of the surrounding air have to be controlled better than those of open atmospheric environments to approach the thermodynamical noise limit of the WGM resonator.

The pressure variation not only changes the ambient air refractive index but could also deform the resonator. Therefore, we consider the resonant frequency shift due to the change in the ambient pressure at the given volume of the WGM resonator leading to a frequency shift by

$$\frac{\Delta \nu_{TM}}{\nu_0} = \frac{1}{3}\beta_T \Delta P, \qquad (4)$$

where $\beta_T = -[(1/V)(\delta V/\delta P)]_T$ is the compressibility of the resonator host material (we assume that isothermal and adiabatic compressibility are approximately equal). The compressibility of MgF$_2$ is $\beta_T = 10^{-11}$ Pa$^{-1}$ and therefore, the frequency shift due to the pressure change ($\Delta P$) of 45 mPa is 29 Hz, which is insignificant in this measurement.

**Frequency noise spectrum of the thermal-compensation WGM resonator**

To minimize the impact of the technical noise from our laboratory environment, the vacuum chamber is evacuated to the pressure of 1.33 mPa but the chamber temperature is not controlled. Figure 3a shows the comparative frequency noise power spectral density (FNPSD) curves of the laser stabilized to the thermal-compensation WGM resonator. For comparison, the theoretically estimated thermodynamical noise limit of the WGM resonator is also plotted. The FNPSD of the free-running laser is shown by the black curve and the noise is substantially suppressed (30 dB at 10 Hz) as shown by the olive curve (b) when the laser is stabilized to the WGM resonator. Below 30 Hz offset frequency, the stabilized laser FNPSD curve falls off as $f^{-1.5}$ implying the thermorefractive noise limit of the many thermal modes of the WGM resonator predicted by Matsko, A. B. *et al.* (Supplementary Note 4). From 30 Hz to 100 Hz, the FNPSD curve falls off with $f^{-1}$ implying the impact of flicker noise caused by the residual laser noise and electronic device noise. Two strong peaks originate from 60 Hz harmonics of the electrical power line noise. The rising frequency noise above 1 kHz in the olive curve is due to the poles in the active feedback loops.

The highest spectral purity laser is achieved among the WGM resonators of the given size and morphology via the thermal-compensation. The integral linewidth is evaluated from the FNPSD and turns out to be 119 Hz but when the two 60 Hz harmonic peaks are removed, it is less than 25 Hz (Supplementary Note 5). To further support our linewidth estimations, we measure the beat signal linewidth (Figure 3b) in a spectrum analyzer with a resolution bandwidth of 47 Hz and sweep time of 40 ms over the 5 kHz span. Due to the slow frequency drift, the measured line shape shows the asymmetry in the wings but the peak center is nearly symmetric and can be fitted with a Lorentzian line shape. The resulting full-width at half-maximum linewidth is approximately 100 Hz, matched with the FNPSD measurement including the 60 Hz noise harmonic peaks.

**Frequency stability of the thermal-compensation WGM resonator**

The resonant frequency instability of the compensated WGM resonator is analyzed by its Allan deviation. We use the FNPSD measurements to evaluate Allan deviations below 0.1 s averaging time because the frequency error of our counter (Agilent 53131A) increases at smaller than 0.1 s averaging times due to the increasing dead time (Allan deviation at 0.1 s averaging time calculated from the FNPSD curve is compared with the values measured by the frequency counter – both are within the error bars of each other). Figure 4a shows the measured Allan deviations. We plot the Allan deviations from 0.01 to 0.1 s averaging time by statistically evaluating 74 measurement traces of FNPSD (circles in red). The lowest measurement has an Allan deviation of 32 Hz at 0.1 s averaging time which agrees with the frequency instability imposed by the theoretically estimated limit from the thermorefractive noise of the WGM resonator. We observe the noise floor at 500 μs averaging time and the Allan deviations show a $\tau^{0.5}$ dependence (magenta line) along the averaging time implying random walk frequency noise, and thus it is different from the theoretically calculated thermorefractive-noise limit (blue curve) excluding the random walk noise statistics.

The statistically estimated Allan deviation at 0.1 s averaging time has a mean value of 95 Hz and a standard deviation of 68 Hz. To understand this deviation, we analyze the mean and standard deviation of 74 FNPSD measurement traces in the low Fourier frequency regime (1 to 40 Hz) shown in Figure 4b. The red line is the mean value and the cyan area is the connected standard deviations along the offset frequency. Both mean frequency noise and standard deviation are diminished along the offset frequency until 20 Hz and reaches the thermorefractive noise limit. However, the residual thermo-mechanical fluctuations still exist near the carrier and change in each measurement causing the standard deviation in Allan deviation measurements. This measurement shows that the WGM resonator has relatively high noise sensitivity at low Fourier offset frequencies and therefore further reduction of the thermal sensitivity via the thermal compensation is desirable to reduce the measurement uncertainty.

**Random walk noise distribution and long-term frequency instability**

The beat frequency at 1 s averaging time is recorded by a frequency counter and the Allan deviations at the longer averaging time are calculated by averaging the 1 s measurement data sets. To check the reproducibility of the stability measurements, we take 10 measurement sets at 1 s averaging time for approximately 10 minutes each and analyze them statistically as illustrated in Figure 4a. The Allan deviations (squares in red) along the averaging time start to deviate from the $\tau^{0.5}$ curve, which implies a frequency drift. To understand this behaviour, we record the temperature of the 10 kΩ thermistor sensor used for detecting the WGM resonator temperature for two and a half hours while the laser is stabilized to the WGM resonator (Figure 5a) and the temperature data is statistically analyzed as shown in Figure 5b. Due to the digitization of the measured temperature, the resolution is limited such that we can only measure the upper and lower temperature bounds. The inset of Figure 5a illustrates that the laser frequency shift is less than 2 MHz during this measurement period and therefore the WGM resonator temperature is actually stabilized with less than 10 mK instability when the measured thermal-expansion sensitivity of 3 ppm $K^{-1}$ is considered. To illuminate the monotonic frequency drift, we apply the random walk and binomial distribution for 30 sets of the number of upper ($n_u$) and lower bound ($n_d$) temperature data points with the number of samples ($N$) from the continuously measured temperature data set (Supplementary Note 6). The different numbers of $N$ are chosen and the average positions ($\overline{m}$) and the standard deviations ($\overline{(\Delta m)^2}$) are calculated for each case. Then we calculate the probability distribution ($P_N(m)$) defined by

$$P_N(m) = \frac{N!}{[(N+m)/2]![(N-m)/2]!} p^{(N+m)/2} q^{(N-m)/2}, \tag{5}$$

where $m=n_u-n_d=2n_u-N$, $p$ ($q$) is the probability that the measured temperature point is at the upper (lower) bound respectively, and they satisfy the relation, $p+q=1$. From the measurement data, we deduce the values of $p$ and $q$ and they are 0.45 and 0.55 respectively. Figure 5b shows the probability distribution for $N=100$, $P_{N=100}(m)$. The average position, $\overline{m}$ is not at the center but at -10.05 and the standard deviation, $\overline{(\Delta m)^2}$ is 9.97 that is approximately $\sqrt{N}$ demonstrating that this measurement follows the random walk distribution. By increasing $N$, the average position shifts monotonically far away from the centre in Figure 5b inset implying that the measured temperature is monotonically shifted to one of the temperature bounds along the increasing integration time, leading to the frequency drift. This frequency shift could be mitigated by improving the resolution of temperature sensing, which is currently limited by the 10 k$\Omega$ thermistor sensor, because the narrower bound of the measured temperature reduces the range of the frequency drift. This could be realized by implementing the dual-mode temperature compensation technique to the WGM resonator [27,28] allowing a detection sensitivity of 100 nK. We also note that the monotonic long-term frequency drift shows a correlation with the monotonic ambient temperature change inferring that the ambient temperature change triggers the feedback to compensate the WGM resonator temperature accordingly. Therefore, an ambient temperature-controlled enclosure might be necessary to enhance the long-term stability.

**Discussion**

We have shown that the impact of the surrounding medium perturbations is a major problem when the WGM resonator stability approaches the thermodynamical fluctuation limit. The ambient temperature stability to reach the thermorefractive noise limit of our thermal-compensation WGM resonator is 2.7 mK and therefore, the environment temperature control is necessary to enhance the frequency stability. An evacuated environment provided such temperature stability in the short integration time such that the thermal-compensation WGM resonator showed the $f^{-1.5}$ frequency noise dependence imposed by thermorefractive noise of the resonator, and allowed the laser linewidth of < 25 Hz and the lowest Allan deviation of 32 Hz in 100 ms integration time, corresponding to the frequency instability of $1.67 \times 10^{-13}$ for the 191 THz carrier. The standard deviation of the Allan deviation measurements is attributed to the thermo-mechanical noise concentrated near the carrier originating from the residual thermal expansion noise, which suggests that the enhanced thermal compensation is desirable to improve the stability. In principle, it is possible to achieve an order of magnitude less thermal-expansion sensitivity which could even more alleviate the required ambient temperature stability. In the longer integration time, the laser stabilized to the WGM resonator experiences the monotonic frequency drift due to the centre-shifted random walk probability distribution of the WGM resonator temperature, which is attributed to the limited temperature sensing accuracy and the correlation of the WGM resonator temperature change with the ambient temperature variation. Hence, we anticipate that the long-term frequency drift could be mitigated by implementing the dual-mode temperature compensation technique and by tightly controlling the environmental temperature. Finally, it is noteworthy that ultrahigh-$Q$ WGM resonators provide enhanced nonlinearity permitting generation of optical frequency combs with low input power, and therefore stable microcombs excluding external references are potentially possible via a single WGM resonator, which could advance the microcomb system in size-, weight-, and power-constrained environments.

**Methods**

**Laser stabilization to the thermal-compensation WGM resonator.** The 15 mW output power from the self-injection locked laser diode is split into two paths by a 50/50 splitter after an acousto-optic modulator (AOM). One arm is used for stabilizing the laser and the other is reserved for characterizing the noise of the resonator by heterodyne-beating the stabilized laser against a 1 Hz FP cavity reference laser. The piezoelectric transducer (PZT) attached to the laser is used to control the laser frequency. The PZT can be controlled either by a digitized signal on computer interface or by an analog voltage input for frequency modulation to obtain the beat frequency within the photodetector bandwidth (New Focus model 1611). The beat note is usually generated between 1 and 1.5 GHz and is down-mixed to 50-100 MHz for counting the beat frequency. We place an AOM before the WGM resonator, which assists the frequency stabilization and advances noise suppression in the acoustic offset frequency regime by extending the feedback bandwidth (Supplementary Note 3). We apply the PDH locking technique to stabilize the laser to the WGM resonator. The laser light (1 mW) is phase-modulated by a fibre-coupled electro-optic modulator (5.5 dB loss) and then launched into the resonator. The transmitted light is detected by an internal photodetector in the packaged aluminum box and the detected signal is demodulated with the same microwave source that phase-modulates the laser light at a double balanced frequency mixer (model: ZAD-1+), which produces an error signal. The error signal is optimized by choosing the optimum modulation frequency and the light intensity into the resonator which are typically ~ 12 MHz and ~ 20 µW respectively. The error signal is split into two branches and fed into commercial high-speed proportional-integral servo controllers (New Focus, LB1005). A slow servo branch acts on the laser PZT and a fast servo branch is used to further suppress acoustic and laser technical noises. The feedback signal from the fast servo controller is fed into a voltage controlled oscillator (VCO). The VCO output is amplified and then applied to the AOM, which shifts the laser frequency to control the frequency noise up to 400 kHz. The contribution of each servo loop is optimized to achieve the lowest noise level.

**Data availability**

Correspondence and requests for materials should be addressed to J.L., A.B.M. and C.W.W.

**References**

[1] Young, B. C., Cruz, F. C., Itano, W. M. & Bergquist, J. C. Visible laser with subhertz linewidths. *Phys. Rev. Lett.* **82**, 3799-3802 (1999).

[2] Parthey, C. G. *et al*. Improved measurement of the hydrogen 1S-2S transition frequency. *Phys. Rev. Lett.* **107**, 203001 (2011).

[3] Takamoto, M., Hong, F.–L., Higashi, R. & Katori, H. An optical lattice clock. *Nature* **435**, 321-324 (2005).

[4] Hollberg, L.*et al*. Optical frequency/wavelength reference. *J. Phys. B: At. Mol. Opt. Phys.* **38** S469-S495 (2005).

[5] Nicholson, T. L. *et al*. Comparison of two independent Sr optical clocks with $1\times10^{-17}$ stability at $10^3$ s. *Phys. Rev. Lett.* **109**, 230801 (2012).

[6] Abbott B. P. *et al*., LIGO: the laser interferometer gravitational-wave observatory. *Rep. Prog. Phys.* **72**, 076901 (2009).

[7] Fortier, T. M. *et al.* Generation of ultrastable microwaves via optical frequency division. *Nat. Photon.* **5**, 425-429 (2011).

[8] Coddington, I., Swann, W. C., Nenadovic, L. & Newbury, N. R. Rapid and precise absolute distance measurements at long range. *Nat. Photon.* **3**, 351-356 (2009).

[9] Thorpe, M. J., Rippe, L., Fortier, T. M., Kirchner, M. S. & Rosenband, T. Frequency stabilization to $6\times10^{-16}$ via spectral-hole burning. *Nat. Photon*. **5**, 688-693 (2011).

[10] Notcutt, M., Ma, L. S., Ye, J. & Hall, J. L. Simple and compact 1-Hz laser system via an improved mounting configuration of a reference cavity. *Opt. Lett.* **30**, 1815-1817 (2005).

[11] Kessler, T. *et al.* A sub-40-mHz-linewidth laser based on a silicon single-crystal optical cavity. *Nat. Photon.* **6**, 687-692 (2012).

[12] Numata, K., Kemery, A. & Camp. J. Thermal-noise limit in the frequency stabilization of lasers with rigid cavities. *Phys. Rev. Lett.* **93**, 250602 (2004).

[13] Cole, G. D., Zhang, W., Martin, M. J., Ye, J. & Aspelmeyer, M. Tenfold reduction of Brownian noise in high reflectivity optical coating. *Nat. Photon.* **7**, 644-650 (2013).

[14] Alnis, J., Matveev, A., Kolachevsky, N., Udem, Th. & Hänsch, T. W. Subhertz linewidth diode lasers by stabilization to vibrationally and thermally compensated ultralow-expansion glass Fabry-Pérot cavities. *Phys. Rev. A* **77**, 053809 (2008).

[15] Fox. R. W. Temperature analysis of low-expansion Fabry-Perot cavities. *Opt. Express* **17**, 15023-15031 (2009).

[16] Armani, D. K., Kippenberg, T. J., Spillane, S. M. & Vahala, K. J. Ultra-high-Q toroid microcavity on a chip. *Nature* **421**, 925-928 (2003)

[17] Grudinin I. S. *et al.* Ultrahigh *Q* crystalline microcavities. *Opt. Commun.* **265**, 33-38 (2006)

[18] Vassiliev, V. V. *et al.* Narrow-line-width diode laser with a high-*Q* microresonator. *Opt. Commun.* **158**, 305-312 (1998).

[19] Liang, W. *et al.* Ultralow noise miniature external cavity semiconductor laser. *Nat. Commun.* **6**, 2468 (2015).

[20] Savchenkov, A. A. *et al.* Stabilization of a Kerr comb oscillator. *Opt. Lett.* **38**, 2636-2639 (2013).

[21] Papp, S. B. *et al.* Microresonator frequency comb optical clock. *Optica* **1**, 10-14 (2014)

[22] Gorodetsky, M. L. & Grudinin, I. S. Fundamental thermal fluctuations in microspheres, *J. Opt. Soc. Am. B.* **21**, 697-705 (2004).

[23] Matsko, A. B., Savchenkov, A. A., Yu, N. & Maleki, L. Whispering-gallery-mode resonators as frequency references. I. Fundamental limitations. *J. Opt. Soc. Am. B.* **24**, 1324-1335 (2007)

[24] Savchenkov, A. A., Matsko, A. B., Ilchenko, V. S., Yu, N. & Maleki, L. Whispering-gallery-mode resonators as frequency references. II. Stabilization. J. Opt. Soc. Am. B. 24, 2988-2997 (2007).

[25] Alnis, J. *et al*. Thermal-noise-limited crystalline whispering-gallery-mode resonator for laser stabilization. *Phys. Rev. A* **84**, 011804 (2011).

[26] Lee, H., Suh, M. -G., Chen, T., Li, J., Diddams, S. A. & Vahala, K. J. Spiral resonators for on-chip laser frequency stabilization. *Nat. Commun.* **4**, 2468 (2013)

[27] Strekalov, D. V., Thompson, R. J., Baumgartel, Grudinin, I. S. & Yu, N. Temperature measurement and stabilization in a birefringent whispering gallery mode resonator. *Opt. Express* **19**, 14495-14501 (2011).

[28] Fescenko, I. *et al*. Dual mode temperature compensation technique for laser stabilization to a crystalline whispering gallery mode resonator. *Opt. Express* **20**, 19185-19193 (2012).

[29] Drever, R. *et al*. Laser phase and frequency stabilization using an optical resonator. *Appl Phys. B* **31**, 97-1065 (1983).

[30] Teraoka, I. & Arnold, S. Perturbation approach to resonance shifts of whispering-gallery modes in a dielectric microsphere as a probe of a surrounding medium, *J. Opt. Soc. Am. B.* **20**, 1937-1945 (2003).

**Acknowledgements**

We acknowledge discussions with Shu-Wei Huang, Zhenda Xie, and Tanya Zelevinsky, along with assistance from Parastou Mortazavian. Funding support is provided by Air Force Research Laboratory under contract FA9453-14-M-0090.

**Author contributions**

J.L., C.W.W, A.B.M. and L.M. designed the experiment and J.L. performed the experiment and analyzed the stabilization measurements. A.A.S. designed and developed the thermal-compensation WGM resonator along with the package assembly, E.D. made the self-injection

locked laser, W.L. assembled the reference WGM resonator, D.E. measured the WGM resonator thermal sensitivity and designed the laser control electronics, and V.I. designed the resonator and laser builds. A.B.M and A.A.S. provided the theory, and all authors helped in the manuscript preparation.

**Additional information**

The authors declare no competing financial interests. Supplementary Information accompanies this paper online. Reprints and permission information is available online at http://www.nature.com/naturecommunications

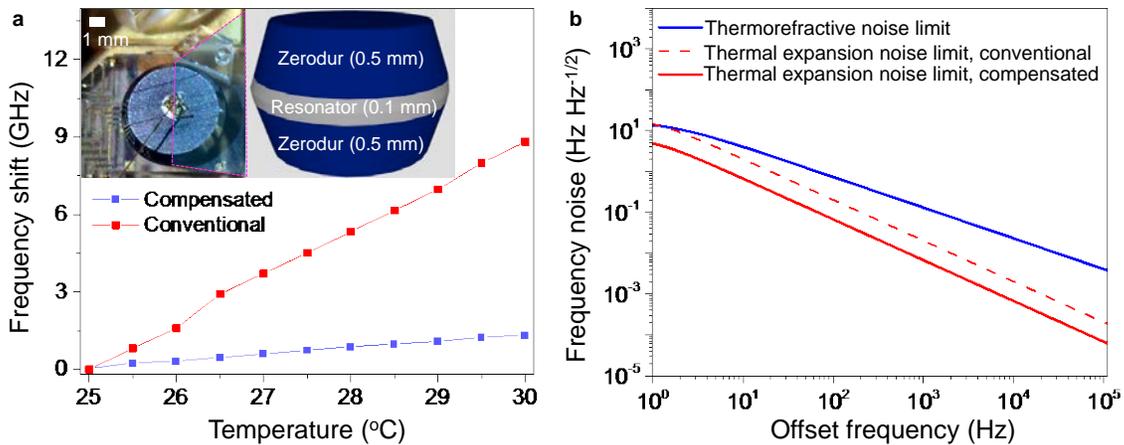

**Figure 1 | Thermal-compensation MgF$_2$ WGM resonator a,** Measurement of the frequency shifts for the conventional MgF$_2$ WGM resonator (red) and thermal-compensation WGM resonator (blue) respectively. Inset: The MgF$_2$ resonator (radius: 3.45 mm and thickness: 0.1 mm) sandwiched by Zerodur layers (thickness: 0.5 mm each) in the packaged unit. The MgF$_2$ resonator (middle) is glued in between two Zerodur components (top and bottom). The thickness of Zerodur layers is determined by the thermo-mechanical properties of the resonator. **b,** Thermorefractive noise limit (blue) and thermal expansion noise limit (red) of the WGM resonator. The simulations show that the thermal expansion noise of the compensated resonator (solid red line) is lower than the thermorefractive noise near the carrier frequency while the conventional one (dashed red line) is similar. Therefore, the fundamental thermorefractive noise fluctuation can be clearly reached with the compensated resonator in this experiment.

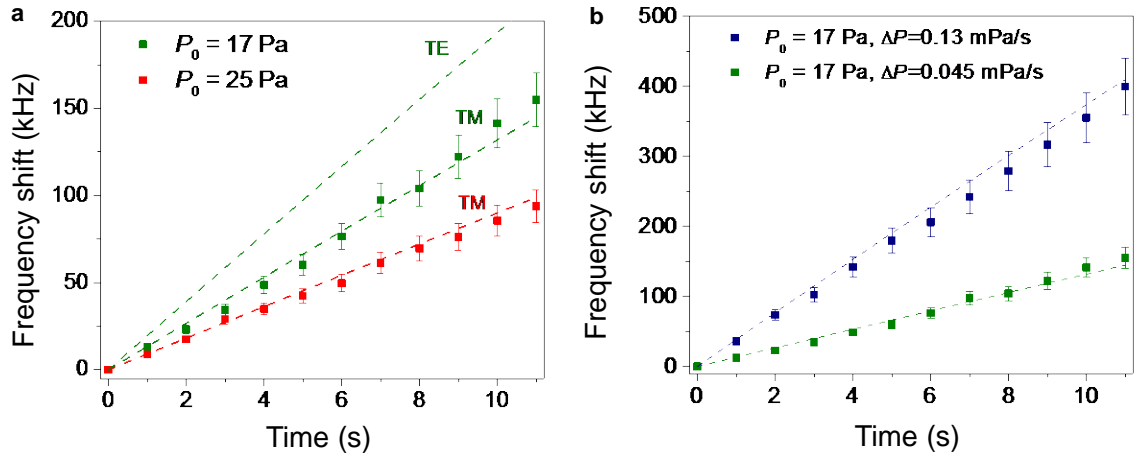

**Figure 2 | Pressure dependence of the WGM resonance frequency shift. a,** The pressure in the vacuum chamber is slowly increased from 17 Pa (olive) and 25 Pa (red) with 45 mPa s$^{-1}$. The frequency shift is measured by counting the beat frequency at every second while the laser is stabilized to the WGM resonator by Pound-Drever-Hall locking. To confirm the frequency shift by the pressure change, the theoretically estimated lines from Eq. (1) for TM and TE modes are plotted together (dashed lines). The measurement values agree with the theoretical model within the 10 % error bar. **b,** The pressure in the vacuum chamber is increased from 17 Pa with 45 mPa s$^{-1}$ (olive) and 130 mPa s$^{-1}$ (navy) respectively. The measurements also agree with the theory within the 10 % error bar.

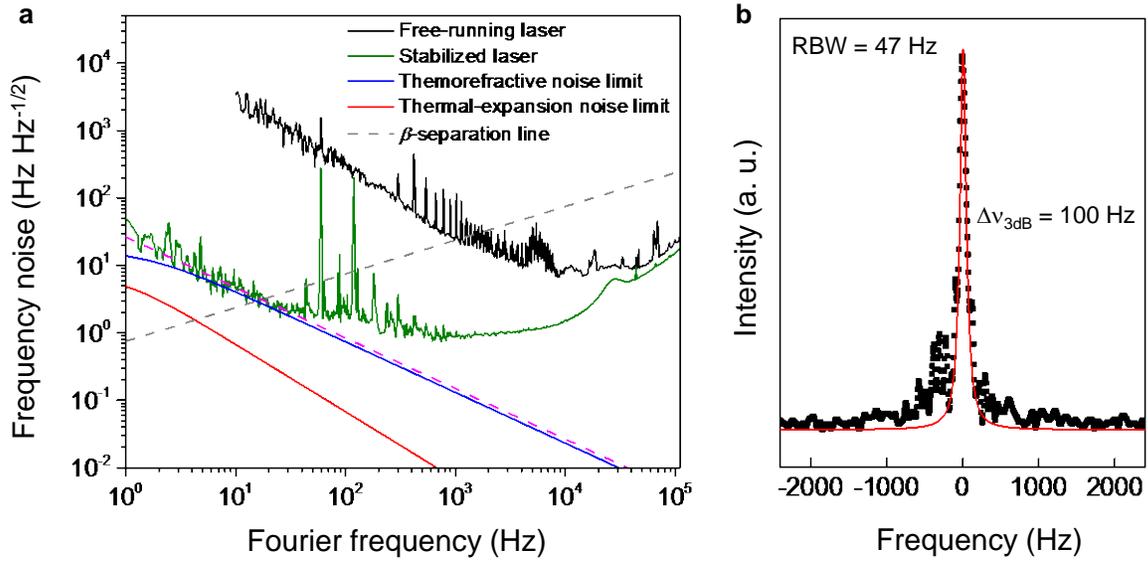

**Figure 3 | Frequency noise of the thermal-compensation WGM resonator. a,** The frequency noise power spectral density (FNPSD) measured by the beat frequency between the laser and the tabletop Fabry-Pérot (FP) cavity reference laser. The FNPSD of free-running laser used in our experiment (black curve) and of the laser stabilized to the compensated resonator (olive curve). The thermodynamical FNPSD predictions from thermorefractive (blue) and thermal expansion noise (red) are also illustrated for comparison. The thermorefractive noise shows $f^{-1.5}$ frequency dependence (magenta dashed line) predicted by our theoretical model. **b,** The measured linewidth (full width at half maximum) on a spectrum analyzer with 47 Hz resolution bandwidth (RBW) and 40 ms sweep time over the 5 kHz span, giving a ≈ 100 Hz Lorentzian linewidth fit. The effective linewidth evaluated from the olive curve is 119 ± 2 Hz and the linewidth without the 60 Hz and 120 Hz power-line noise is determined to be ~ 25 ± 0.3 Hz.

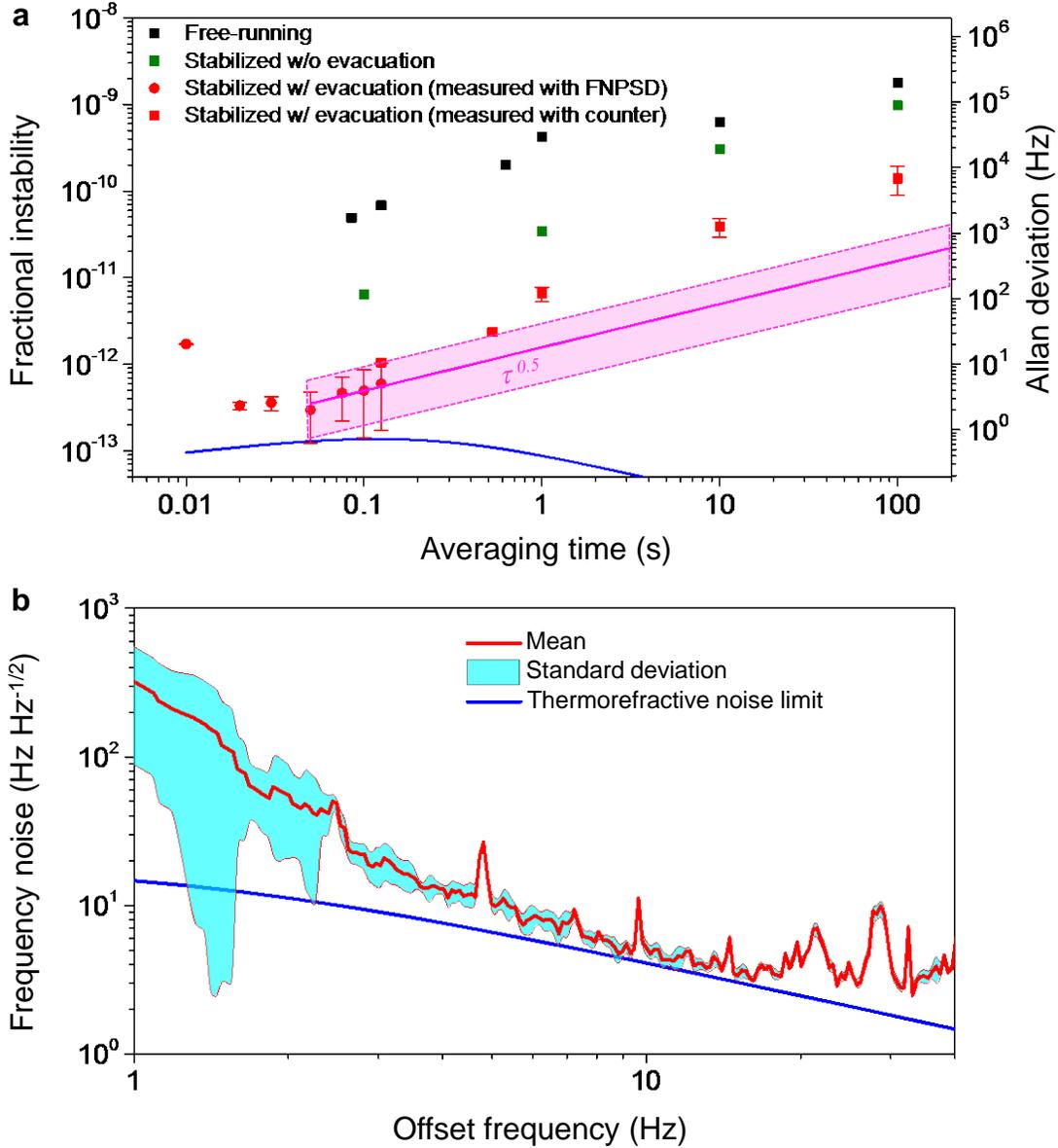

**Figure 4 | Fractional frequency instability measurements. a,** Measured fractional frequency instabilities of the beat frequency between the Fabry-Pérot (FP) cavity reference laser and the free-running laser (squares in black), and the beat frequency between the reference laser and the laser stabilized to the WGM resonator without evacuation (squares in olive) respectively. The red squares and bars are the mean ($\mu$) and standard deviation ($\sigma$) of the frequency instability of the laser stabilized to the resonator calculated from 10 sets of frequency counting data at 1 s averaging time respectively, measured with a frequency counter. The red circles and bars are $\mu$ and $\sigma$ of the frequency instability derived from the 74 frequency noise power spectral density (FNPSD)

measurement traces. The magenta area represents the expected Allan deviation bound imposed by the random walk frequency noise. **b,** The noise statistics of 74 FNPSD traces. The red line is the mean value and the cyan area is the connected standard deviations. The blue line is the thermorefractive noise limit. Both mean frequency noise and standard deviation are reduced along the offset frequency and the frequency noise reaches the thermorefractive noise limit at 6 Hz. However, the thermo-mechanical noise, induced by the thermal expansion, is still concentrated near the carrier causing the standard deviation of the Allan deviation measurements in Figure 4(a).

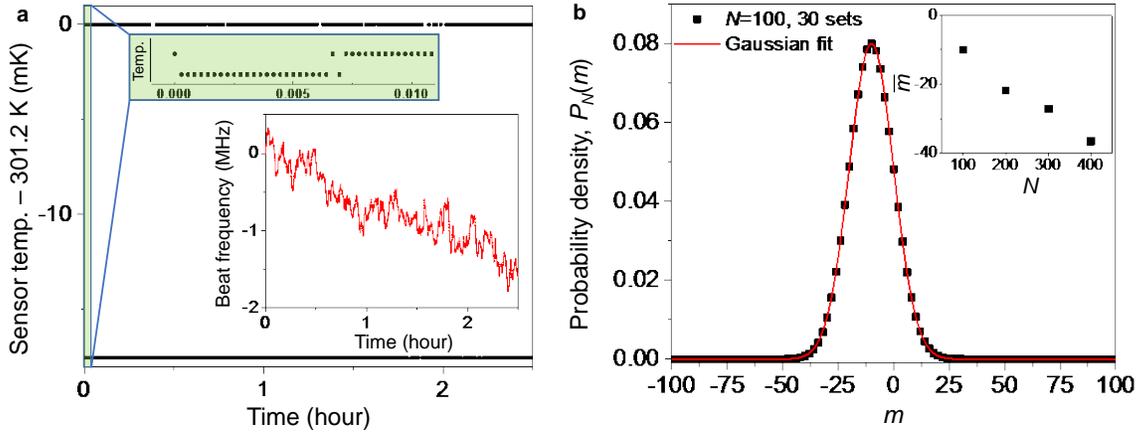

**Figure 5 | Thermistor sensor temperature random walk distribution statistics. a,** The 10 kΩ thermistor sensor temperature measurement data. Due to the temperature digitization, the measurement resolution is limited such that the upper and lower temperature bounds per every one second are recorded. Inset: resonance frequency shift during the temperature measurement period. **b,** The probability distribution for $N=100$, $P_{N=100}(m)$, where $\overline{m}$ is located at -10.05 and $\overline{(\Delta m)^2}$ is 9.97, which is approximately $\sqrt{N}$ demonstrating that this measurement follows the random walk distribution. The inset shows that the increased number of samples monotonically shifts $\overline{m}$ far from the center.

**Supplementary Note 1. Fabrication of a thermal-compensation whispering-gallery-mode resonator**

To compensate for the extensive thermal expansion of a whispering-gallery-mode (WGM) resonator, the resonator is sandwiched by a nearly zero or slightly negative expansion material (e.g. Zerodur). The compensating layers should be thick enough to hold the resonator's expansion. The ambient temperature variations should be small ($< 5\ ^{\circ}C$) to avoid resonator destruction. The layers are bonded together with a transparent glue and then annealed with slightly varying temperature.

We determined the thickness of the compensating layer on the resonator to fully suppress an overall frequency shift for a given resonator's diameter and thickness via COMSOL simulations (Supplementary Figure 1a). The glue layer has 2 μm thickness and is taken into account in the numerical simulation. The colour code indicates the linear expansion of the WGM resonator's equator. The blue colour represents negative expansion dominated by Zerodur components and the red color represents positive expansion where Zerodur components are too weak to hold the resonator's expansion. The green zone shows zero expansion where Zerodur contraction and $MgF_2$ expansion compensate each other. The thermal-expansion coefficient of the crystalline $MgF_2$ is approximately 9 ppm $K^{-1}$ in the direction that is orthogonal to the crystalline/resonator axis. Based on the simulations, we found that the coefficient can be reduced to 0.15 ppm $K^{-1}$ (60 times) by choosing the proper thickness of the Zerodur layers in the resonator structure described in the inset of Figure 1a unless the stress that propagates to the localized WGM takes into account. We should note here that all other sources of drift such as mechanical creeps or fast gradients of the temperature across the resonator are not eliminated by this technique.

For the compensated resonator, the temperature-dependent resonance frequency shift is introduced by three factors. One of them is the change in the thermorefractive index, which is small (0.6 ppm $K^{-1}$) for the ordinary axis in $MgF_2$ crystal. The other two factors are the residual thermal expansion and elasto-optical refractive index changes caused by the stress induced in the material when its expansion is prevented by the Zerodur layers. By using the data for the thermorefractive effect obtained from the Corning $MgF_2$ datasheet and data for the elasto-optical effect in ref. [1], we evaluated the stress-induced refractive index change value, averaged over the mode volume and found 0.25-0.5 ppm $K^{-1}$ (the variation is attributed to the uncontrollable changes of the glue layer), which is smaller than the thermorefractive index change (0.6 ppm $K^{-1}$). The reduction of the impact of the ambient temperature fluctuations on the resonator enables the measurement of the fundamental thermorefractive noise we performed.

The compensation of the free standing resonator may be affected by its mounting technique in a particular resonator's housing. To make sure that our resonator is still compensated after installation, we simulated the expansion of components with an entire mechanical model (Supplementary Figure 1b). Although the complete compensation is lost when the resonator is mounted into the housing, the predicted suppression of the resonator's expansion remains 23× - an acceptable value for this experiment.

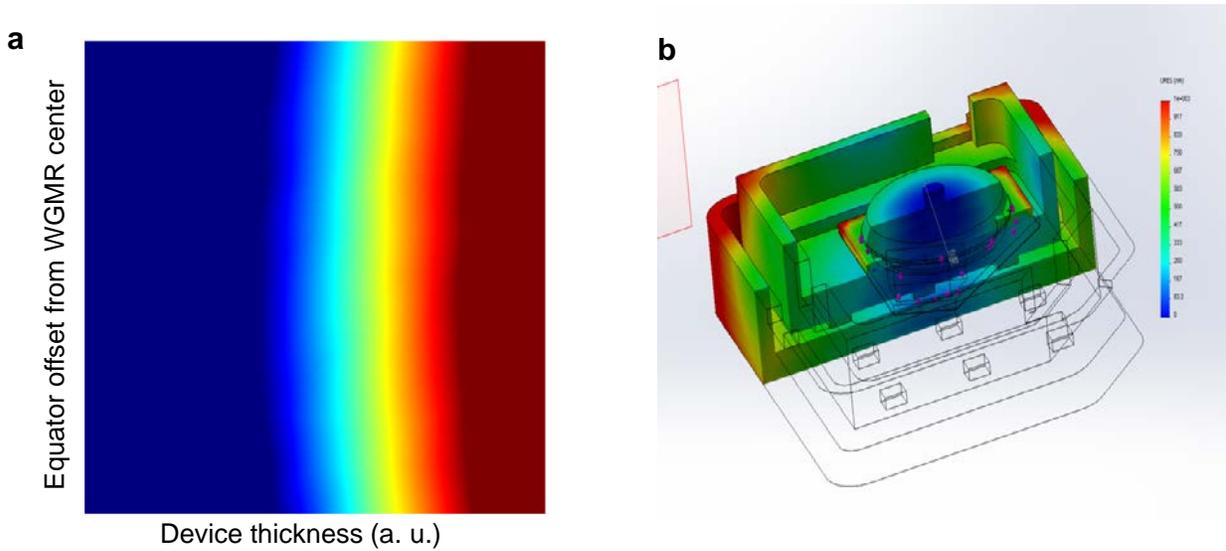

**Supplementary Figure 1 | Simulations of the thermal-compensation. a,** The simulation of the whispering-gallery-mode (WGM) resonator expansion depending on the device thickness. The colour code indicates the linear expansion of the WGM resonator's equator. The blue colour zone represents negative expansion largely governed by Zerodur components and the red colour zone represents positive expansion where the $MgF_2$ resonator's expansion is dominant. The green colour

zone shows zero expansion where Zerodur contraction and MgF$_2$ expansion compensate each other. **b,** A simulation of thermal expansion of the device in the package mount.

**Supplementary Note 2. Thermodynamical noise limits of the cylindrical WGM resonator**

The Supplementary Figure 2a illustrates the WGM resonator used in this experiment and its ring-down time (4 μs) shows the loaded $Q$ of 2.4×10$^9$. The WGM resonator has the radius ($r$) of 3.45 mm, the rim thickness ($L$) of 0.025 mm. The thermodynamical noise limit of the compensated MgF$_2$ WGM resonator is calculated by our theoretical model [2,3]. Here, we consider the two major thermal noise sources that are the thermal-expansion noise and the thermorefractive noise. We numerically quantify the frequency noise power spectral density (FNPSD) imposed by thermal expansion noise for comparing it with experimental measurements using

$$S_\nu^2(f)_{tex} = \nu_0^2 \frac{k_B \alpha_l^2 T^2}{\rho C V_c} \frac{2r^2/\pi^2 D}{1+\left(2fr^2/D\pi\right)^2}, \tag{1}$$

where $\alpha_l$ is the thermal expansion coefficient and $V_c$ is the volume of the WGM resonator. The thermorefractive noise limit is also evaluated by Eq. 2, which is derived for the cylindrical WGM resonator possessing $r \gg L$.

$$S_\nu^2(f)_{tre} = \nu_0^2 \frac{k_B \alpha_n^2 T^2}{\rho C V_m} \frac{r^2}{12D} \left[1+\left(\frac{2\pi r^2 |f|}{9\sqrt{3}D}\right)^{3/2} + \frac{1}{6}\left(\frac{r^2}{D}\frac{\pi f}{4m^{1/3}}\right)^2\right]^{-1}, \tag{2}$$

where $\nu_0$ is the carrier frequency, $k_B$ is Boltzmann constant, $\alpha_n$ is the thermorefractive coefficient of the material, $\rho$ is the material density, $C$ is the specific heat capacity, $V_m$ is the mode volume, $D$ is the heat diffusion coefficient, and $m$ is the mode order defined by $m=2\pi r n \lambda^{-1}$. This equation shows the $f^{-1.5}$ frequency dependence for the compensated WGM resonator. The values of the parameters used in these simulations are $\alpha_l$ = 9×10$^{-6}$ K$^{-1}$ (3×10$^{-6}$ K$^{-1}$) for the conventional (compensated) WGM resonator, $\alpha_n$ = 6×10$^{-7}$ K$^{-1}$, $T$ = 300 K, $\rho$ = 3.18 g·cm$^{-3}$, $C$ = 9.2×10$^6$ erg·g$^{-1}$·K$^{-1}$, $V_m$ = 2.62×10$^{-6}$ cm$^3$, $D$ = 7.17×10$^{-2}$ cm$^2$·s$^{-1}$, $n$ = 1.37, $V_r$ = 9.34×10$^{-4}$ cm$^3$ and $\lambda$ = 1565.5 nm. We simulate the equations and compare the noise of the thermal-compensation WGM resonator with that of the conventional WGM resonator as shown in Figure 1b, which shows the thermal compensation reduces the thermal expansion noise below the thermorefractive noise near the carrier frequency. Here, we also experimentally demonstrate the reduction of the thermal expansion noise for the compensated WGM resonator by stabilizing a laser to it. By comparing the FNPSD of the compensated resonator with that of a conventional resonator, possessing approximately equal quality factors, we observe the reduction of frequency noise near the carrier frequency from 1 to 100 Hz by greater than 3 dB as shown in Supplementary Figure 2b.

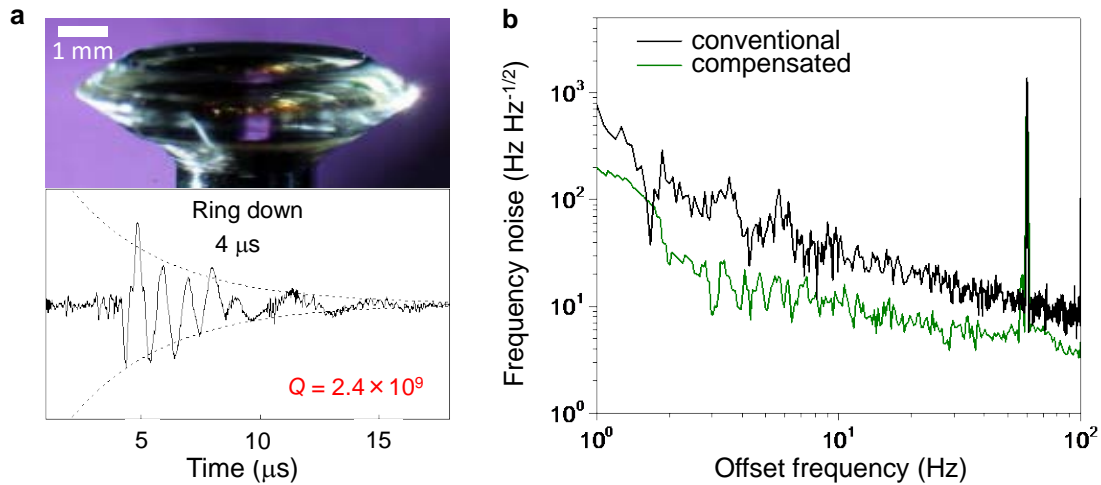

**Supplementary Figure 2 | Compensated WGM resonator characteristics. a,** The whispering-gallery-mode (WGM) resonator has 6.9 mm in diameter and 0.1 mm in height. The thickness of rim is 25 μm. The ring-down measurement of the resonator shows 4 μs ring-down time corresponding to $Q$ of $2.4\times10^9$. **b,** Experimentally measured frequency noise of the beat frequency between the WGM resonator-stabilized laser and 1 Hz reference laser for the compensated (olive) and the conventional (black) WGM resonator respectively. The noise is reduced by greater than 3 dB from 1 Hz to 100 Hz offset frequencies.

**Supplementary Note 3. Pound-Drever-Hall laser locking and noise characterization**

Supplementary Figure 3a illustrates the laser stabilization and noise characterization set-up. A compact 3 kHz linewidth microlaser is coupled into a thermal-compensation WGM resonator in a rigid vacuum chamber and stabilized via Pound-Drever-Hall (PDH) locking technique with both a piezoelectric transducer (PZT) and an acoustic-optical modulator (AOM). The FNPSD and relative frequency instability of the stabilized microlaser are analyzed by heterodyne-beating against a Fabry-Pérot (FP) cavity reference laser, possessing 1 Hz linewidth and less than 0.1 Hz s$^{-1}$ drift-rate. The PDH error signal is optimized for the best laser-WGM resonator stabilization. We modulate our laser frequency by applying a ramping ($f_{mod}$ = 18 Hz) voltage signal to either the piezoelectric transducer (PZT) in the laser or the AOM to search a single WGM resonance near the reference laser frequency (191 THz). The transfer function for laser frequency shift per modulation voltage is 4.5 MHz V$^{-1}$. Supplementary Figure 3b shows the error signals during the periodic ramping modulation within a half period of time. We carefully optimize the input optical power and polarization to the WGM resonator and phase modulation frequency and amplitude to an electro-optic modulator to find a single error signal (blue curve), which provides tight and stable laser frequency stabilization. Using the weaker error signal from the two resonances (red curve) within the modulation period, the laser frequency stabilization is less stable.

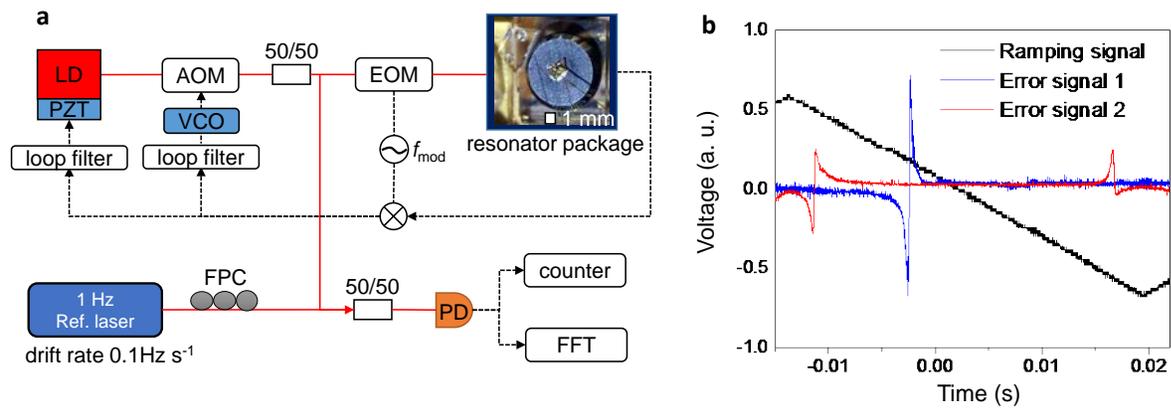

**Supplementary Figure 3 | Laser frequency stabilization and noise characterization. a,** A microlaser with 3 kHz linewidth is stabilized to the reference whispering-gallery-mode (WGM) resonator via Pound-Drever-Hall (PDH) locking technique. The piezoelectric transducer (PZT) attached onto the microlaser and the external acoustic-optical modulator (AOM) are used to minimize technical noise such that the laser frequency chases a WGM resonance frequency**.** The resonance frequency shift is measured by beating the WGM resonator-stabilized laser frequency

against a reference laser possessing 1 Hz linewidth and 0.1 Hz s$^{-1}$ drift rate. LD: laser diode; EOM: electro-optic modulator; PD: photodetector; FPC: fiber polarization controller. **b,** The laser frequency is modulated to search a WGM resonance near the reference laser frequency (191 THz). The ramping modulation voltage with the frequency of 18 Hz is applied to the laser and the laser frequency shift per modulation voltage is 4.5 MHz V$^{-1}$. The two different error signals corresponding to different resonances are used for stabilizing the laser to the WGM resonator. The tight and stable laser stabilization is only achieved with the strong single error signal (blue curve) within the modulation period.

Supplementary Figure 4a shows the FNPSD curves for two different ambient pressure levels. We observe many noise spikes in the acoustic frequency regime originating from the laboratory environment when the WGM resonator is just sealed without evacuating the vacuum chamber (black curve). These spikes are suppressed by increasing the vacuum level to 1.33 mPa (red curve). The remaining spikes mostly come from the 60 Hz harmonics of the electrical power-line. In this measurement, we observed that the evacuated chamber substantially reduces acoustic noise peaks and provides further noise reduction near the carrier frequency. Since the PZT control loop has a limited feedback bandwidth, we add the second feedback loop using an external AOM to increase the feedback bandwidth for further technical noise suppression. With the AOM feedback loop, the feedback bandwidth is extended over 400 kHz. As a result, the noise over the 20 Hz offset frequencies is far more suppressed than those with only PZT feedback control (Supplementary Figure 4b), which allows us to investigate the thermodynamical noise limit of the thermal-compensation WGM resonator more clearly.

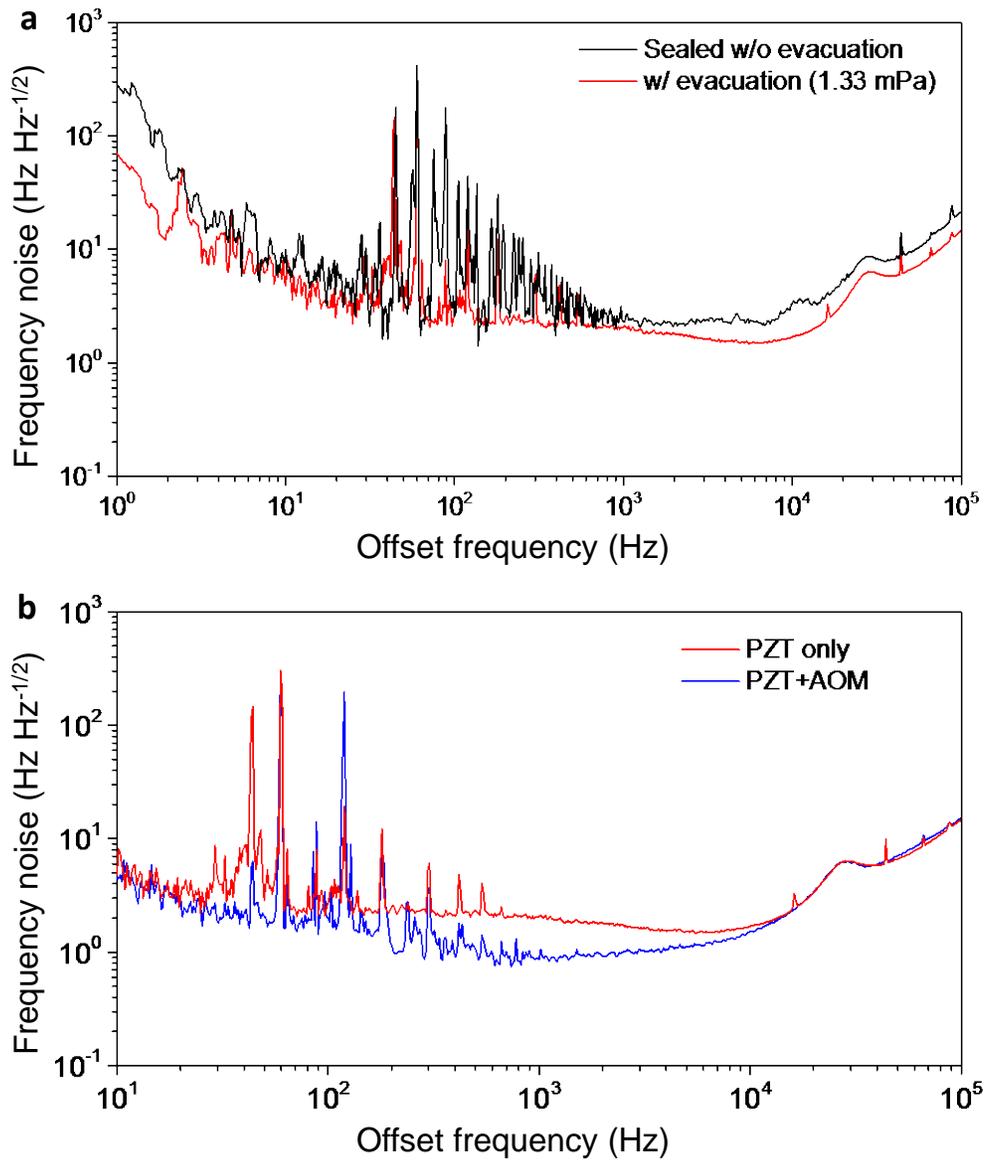

**Supplementary Figure 4 | Impacts of the evacuation and the external AOM feedback loop. a,** The frequency noise power spectral density (FNPSD) in a sealed chamber without evacuation (black curve). Many spikes are observed in the acoustic noise frequency regime. The FNPSD with evacuation (red curve) in which its pressure is 1.33 mPa. Most of the noise spikes are mitigated and the remaining spikes are mostly from the 60 Hz harmonics of the electrical power-line. **b,** The

FNPSD with PZT only feedback loop (red curve), and the FNPSD with both PZT and acosto-optic modulator (AOM) feedback loops (blue curve).

**Supplementary Note 4. Why thermorefractive noise of a WGM resonator is pink**

In our work we reached the fundamental thermodynamical noise floor of the resonator and found out that the power spectrum of the noise scales as $f^{-1.5}$. This is so called pink or flicker noise which is general for various physical processes, ranging from traffic flow to DNA sequence structure [4]. Flicker noise is the major physical mechanism that limits spectral purity of lasers stabilized to super-cavities [5]. The origin of this noise is still not completely understood and is usually studied case-specifically [4-6]. It was theoretically predicted that the fundamental thermorefractive frequency noise of a WGM resonator has distinct pink noise dependence [7]. In this work, we provide a careful experimental study of this noise.

The pink noise in the WGM resonator is a direct consequence of the presence of multiple normal modes of the heat transfer equation in the WGM resonator. The power spectral density of the thermal distribution can be found by summing over these normal modes of a thin resonator of radius $R$ and thickness $L$ ($R \gg L$) [8]

$$S(\Omega) = \frac{k_B T^2}{\rho C V_m} \frac{16D}{R^2} \sum_{p \leq \nu^{2/3}} \sum_{m=0,2} \frac{1}{\Omega^2 + D^2 k_{p,m}^4} \approx \frac{k_B T^2}{\rho C V_m} \frac{R^2}{12D} \left[ 1 + \left( \frac{R^2}{D} \frac{|\Omega|}{9\sqrt{3}} \right)^{3/2} + \frac{1}{6} \left( \frac{R^2}{D} \frac{\Omega}{8\nu^{1/3}} \right)^2 \right]^{-1} \quad (3)$$

where $\Omega = 2\pi f$ is the spectral frequency, $k_B$ is the Boltzmann constant, $T$ is the ambient temperature, $\rho$ is the density of the resonator host material, $C$ is the heat capacity of the material, $V_m$ is the volume of the resonator mode, $D$ is the heat diffusion coefficient of the resonator host material, $\nu$ is the azimuthal number of the optical WGM, and $k_{p,m}$ is the eigenvector of the heat transfer equation. The power density of the frequency fluctuations of WGMs is proportional to $S(\Omega)$ and its proportionality coefficient is *in situ* the squared thermorefractive coefficient of the material, $\alpha_n^2$. From this equation, we find that spectral density of the fundamental thermorefractive noise has Brownian noise frequency dependence ($\Omega^{-2}$) when we only consider one thermal mode of the WGM resonator. However, summing over many modes results in the pink frequency noise ($\Omega^{-1.5}$).

Our experimental data confirms this theoretical prediction. This is a fundamentally important observation for a physical structure with limited dimensionality. It is worth noting that similar results were predicted for optical mirror coatings which is important in classical and quantum metrology [8,9].

**Supplementary Note 5. Sub-100 Hz linewidth WGM resonator-stabilized laser**

Characterizing the spectral linewidth of the cavity-stabilized laser from the FNPSD is not trivial due to the non-unified definition of linewidth of a flickering or drifting laser frequency and therefore we estimate spectral linewidth with two different methods. The integral linewidth is first calculated from the power spectral density with the phase noise method [10]. An effective linewidth of 119 ± 2 Hz is deduced from the raw FNPSD measurement from the olive curve in Figure 3a. We note that most of the residual noise and linewidth contribution arises from the 60 Hz and 120 Hz electrical power line frequency noise. When these two dominant noise peaks are excluded, an upper bound 25 ± 0.3 Hz is estimated. By fitting the frequency noise dependence using frequency decomposition method that removes the rest of the spurious frequencies, we estimate the fundamental noise limited spectral purity and its resultant linewidth is determined to be 8.7 Hz. The spectral linewidth is double-checked by the $\beta$-separation line method [11]. This approach implements a simple geometric line based on the low-pass filtered white noise to determine the laser linewidth for an arbitrary frequency noise spectrum. The $\beta$-separation line is plotted in Figure 3a (dashed gray line). By integrating the FNPSD up to 16 Hz offset frequency which is the region above the $\beta$-separation line, we determine the linewidth of 30.6 ± 0.1 Hz.

**Supplementary Note 6. Random walk noise distribution**

We calculated the probability distribution [12] for $N$=100, where $N$ is the number of samples. The probability distributions for the larger $N$ are also similarly calculated. We took 30 sets of samples of temperature data. Supplementary Table 1 shows the number of upper bound temperature data points ($n_u$) for each set of data. The average of $n_u$ ($\overline{n_u}$) is calculated by $\frac{1}{30}\sum_{s=1}^{30} n_{us}$ and $\bar{m}$ is derived by $2\bar{n_u} - N$. The $p$ and $q$ are computed by solving the coupled equations that are $p + q = 1$ and $p - q = \bar{m}/N$, where $p$ ($q$) is the probability that the measured temperature point is at the upper (lower) temperature bound. The probability distribution for $N=100$ is plotted in Figure 5b and its standard deviation, $\overline{(\Delta m)^2}$ is measured by a Gaussian function fit. The standard deviation is approximately $\sqrt{N}$, implying that this measurement follows the random walk distribution.

| Set # | 1 | 2 | 3 | 4 | 5 | 6 | 7 | 8 | 9 | 10 |
|---|---|---|---|---|---|---|---|---|---|---|
| $n_u$ | 37 | 36 | 53 | 52 | 52 | 43 | 53 | 53 | 48 | 63 |
| Set # | 11 | 12 | 13 | 14 | 15 | 16 | 17 | 18 | 19 | 20 |
| $n_u$ | 37 | 55 | 51 | 11 | 32 | 32 | 37 | 42 | 57 | 77 |
| Set # | 21 | 22 | 23 | 24 | 25 | 26 | 27 | 28 | 29 | 30 |
| $n_u$ | 47 | 38 | 39 | 75 | 19 | 57 | 53 | 51 | 18 | 40 |

**Supplementary Table 1 | The number of upper bound temperature data points ($n_u$) for each set of data.**

**Supplementary References**

[1] Szczesniak, J. P. & Corelli, J. C. Stress optical properties of solids in the 1 to 20 micron wavelength region, Rensselaer Polytechnic Institute, Department of Nuclear Engineering, Troy, NY.

[2] Matsko, A. B., Savchenkov, A. A., Yu, N. & Maleki, L. Whispering-gallery-mode resonators as frequency references. I. Fundamental limitations. *J. Opt. Soc. Am. B.* **24**, 1324-1335 (2007)

[3] Savchenkov, A. A., Matsko, A. B., Ilchenko, V. S., Yu, N. & Maleki, L. Whispering-gallery-mode resonators as frequency references. II. Stabilization. *J. Opt. Soc. Am. B.* **24**, 2988-2997 (2007)

[4] Gilden, D. L., Thornton, T. & Mallon, M. W. 1/*f* noise in human cognition. Science **267**, 1837-1839 (1995)

[5] Kessler, T. *et al*. A sub-40-mHz-linewidth laser based on a silicon single-crystal optical cavity. *Nat. Photon.* **6**, 687-692 (2012)

[6] Hooge, F. N., Kleinpenning, T. G. & Vandamme, L. K. Experimental studies on 1/*f* noise. *Reports on progress in Physics* **44**, 480-532 (1981)

[7] Matsko, A. B., Savchenkov, A. A., Yu, N. & Maleki, L. Whispering-gallery-mode resonators as frequency references. I. Fundamental limitations. *J. Opt. Soc. Am. B.* **24**, 1324-1335 (2007)

[8] Braginsky, V. B. & Vyatchanin, S.P. Thermodynamical fluctuations in optical mirror coatings. *Physics Letters A*. **312**, 244-255 (2003).

[9] Evans, M. *et al.* Thermo-optic noise in coated mirrors for high-precision optical measurements. *Phys. Rev. D.* **78**, 102003 (2008)

[10] Hjelme, D. R., Mickelson, A. R. & Beausoleil, R. G. Semiconductor laser stabilized by external optical feedback. *IEEE J. Quantum Electron.* **27**, 352-372 (1991)

[11] Domenico, G. D., Schilt, S. & Thomann P. Simple approach to the relation between laser frequency noise and laser line shape. *App. Opt.* **49**, 4801-4807 (2010)

[12] Reif, F. Fundamentals of statistical and thermal physics. International Edition, chap. **1**, (*McGraw-Hill*, 1985)